\documentclass[aps,groupedaddres8s,fleqn,nofootinbib,twocolumn]{revtex4}
\usepackage{graphicx}
\usepackage{xcolor}
\usepackage{amsmath,amssymb}
\usepackage{float}
\usepackage{physics}
\usepackage{amsfonts}
\usepackage{verbatim}
\usepackage{flushend}
\usepackage{balance}
\usepackage{endnotes}
\usepackage{footnote}
\usepackage{adjustbox}
\usepackage{gensymb}
\usepackage{float}
\usepackage{epigraph}
\usepackage{xcolor}
\usepackage[utf8]{inputenc}
\usepackage{setspace}

\newcommand{\beq}{\begin{eqnarray}}
\newcommand{\eeq}{\end{eqnarray}}
\usepackage{amsmath}

\begin{document}

\title{Constraints on fundamental physical constants from bio-friendly viscosity and diffusion}
\author{K. Trachenko$^{1}$}
\address{$^1$ School of Physical and Chemical Sciences, Queen Mary University of London, Mile End Road, London, E1 4NS, UK}

\begin{abstract}
The problem of understanding fundamental physical constants was discussed in particle physics, astronomy and cosmology. Here, I show that a new insight comes from condensed matter physics and liquid physics in particular: fundamental constants have a bio-friendly window constrained by bio-friendly viscosity and diffusion setting the motion in essential life processes in and across cells. I also show that bounds on viscosity, diffusion and the fundamental velocity gradient in a biochemical machine can all be varied while keeping the fine-structure constant and the proton-to-electron mass ratio intact, with no implication for the production of heavy nuclei in stars. This leads to a conjecture of multiple tuning and an evolutionary mechanism.
\end{abstract}

\maketitle

\section{Introduction}

The origin and values of fundamental physical constants were discussed in high-energy particle physics, cosmology and astronomy \cite{barrow,barrow1,carr,cahnreview,hoganreview,adamsreview,uzanreview,carrbook,finebook,weinberg}. These constants give our world its distinctive character and differentiate it from others we might imagine. The values of these constants have no explanation and are therefore considered arbitrary \cite{cahnreview}, for the reason that we do not know what kind of theories we need to explain them \cite{weinberg}. Understanding fundamental constants is viewed as one of the grandest questions in modern science \cite{grandest}.

The values of some fundamental physical constants are considered to be finely-tuned and balanced to give our observable world. Examples include finely-tuned balance between quark masses needed to produce protons and neutrons \cite{hoganreview,hoganbook,adamsreview} and production of heavy nuclei in stars which depends on the finely-tuned balance between the fine structure constant $\alpha=\frac{e^2}{\hbar c}\approx\frac{1}{137}$ ($e$ is the electron charge and $c$ is the speed of light) and the ratio of the proton mass $m_p$ and electron mass $m_e$, $\beta=\frac{m_p}{m_e}\approx 1836$.
%The existence of the Hoyle resonance levels needed for carbon abundance depends on the finely-tuned balance between $\alpha$ and the strong nuclear force constant \cite{barrow,carrbook,finebook}.
These and other example suggest a narrow ``habitable zone'' in parameter space ($\alpha$,$\beta$) where essential biochemical elements can form \cite{barrow,carrbook,finebook} (see, however, Ref. \cite{adamsreview}). For this reason, fundamental constants are referred to as ``bio-friendly'' or ``biophilic'' \cite{barrow,adamsreview}. Trying to rationalise fundamental constants, their balance and tuning has given rise to the anthropic principle \cite{barrow,barrow1,hoganreview,adamsreview,uzanreview,carr,carrbook,finebook}.
%The weak anthropic principle says that our expectations to observe must be restricted by the condition necessary for our presence as observers. To explain the coincidences such as the Hoyle resonance and fine-tuning of fundamental constants, the strong anthropic principle proposes that the Universe and its fundamental constants must be such as to admit at some stage the emergence of observers \cite{barrow}.

Discussions of constraints on fundamental constants and their fine tuning involve high-energy processes at different scales and often end with production of heavy nuclei in stars. This involves a tacit assumption that once heavy nuclei are produced, observers emerge. However, there are about 15 orders of magnitude size difference between nuclei and observers. Many life processes, including the formation of proteins, RNA, living cells and so on operate on the scale of length and energy considered in condensed matter physics. Due to their complexity and variety, these processes were not thought to be describable by a physical model which can relate them to fundamental constants and put bio-friendly constraints on these constants \cite{barrow1,ellisuzan}. The challenge is to have a physical model which is both general enough to be widely applicable and specific enough to connect life processes directly to fundamental constants.

Here, I show that these models are nevertheless possible. These models are general enough to impose constraints on fundamental constants from bio-friendly viscosity and diffusion involved in essential life processes setting the motion in and across cells. These constraints imply a bio-friendly window for fundamental constants. I show that bounds on viscosity and diffusion can be varied while keeping $\alpha$ and $\beta$ intact, with no implication for the production of heavy nuclei. The same applies to the fundamental velocity gradient which I introduce in relation to flow in a biochemical machine. These observations lead to a conjecture of multiple tuning and an evolutionary mechanism.

\section{Results and Discussion}

\subsection{Motion and flow}

I consider the cell, the basic building block of life forms. There are several areas related to cells where flow is important. The two important ones are the operation of the cell itself (e.g., transport involving protein motors and cytoskeletal filaments, passive and active molecular transport, cytoplasmic mixing, mobility of cytoplasmic constituents, diffusion involved in cell proliferation \cite{cellpaper1,cellpaper2}) and so on) and the flow in the organism involving many cells (e.g., blood flow). Another area where flow is important is related to the pre-biotic synthesis of life building blocks in the metabolic flux, the basis of life, thought to give rise to DNA blocks in protocells \cite{lanebook}. Liquids and gases are two states providing a medium where this flow can happen and matter can {\it move}. Viscosity governs this flow and is therefore tightly embedded in life processes and their dynamics.

In our world, the motion-enabling liquid is water, however the physical model discussed below and its implications apply to all liquids. If life in a different world is not water-based but uses another liquid as a medium to provide motion, the model implications are the same.

I recall that viscosity universally has minima seen in Fig. 1. The minima correspond to the crossover between liquidlike and gaslike dynamics. The data in Fig. 1 are shown above the supercritical pressure in order to extend the temperature range where the system is a fluid. Above the critical point, the minima are smooth and slightly increase with pressure. Below the critical point, viscosity has sharper change at the liquid-gas transition, however viscosity minima above and below the critical point are close \cite{sciadv1}.

The kinematic viscosity $\nu$ at the minimum, $\nu_{min}$, can be evaluated as

\begin{figure}
\begin{center}
{\scalebox{0.35}{\includegraphics{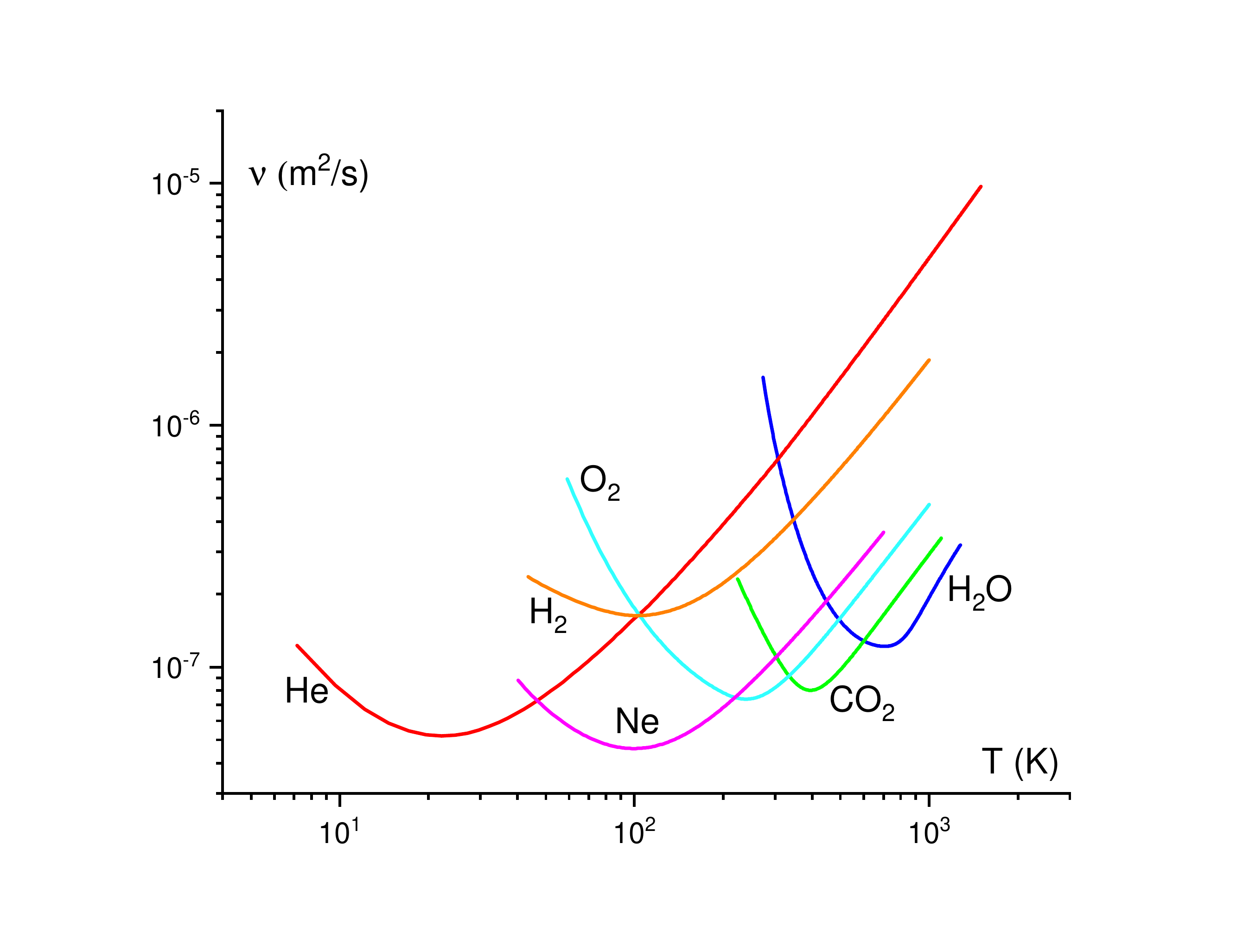}}}
\end{center}
\caption{Experimental kinematic viscosity of noble, molecular and network fluids showing minima. $\nu$ for He, H$_2$, O$_2$, Ne, CO$_2$ and H$_2$O are shown at 20 MPa, 50 MPa, 30 MPa, 50 MPa, 30 MPa and 100 MPa, respectively. The experimental data are from Ref. \cite{nist}.
}
\label{minima}
\end{figure}

\begin{equation}
\nu_{min}=\frac{1}{4\pi}\frac{\hbar}{\sqrt{m_em}}
\label{nu1}
\end{equation}

\noindent where $m$ is the molecule mass and is in agreement with viscosity minima seen in a wide range of experimental data \cite{sciadv1}. The lower viscosity bound \eqref{nu1} is also consistent with the high-temperature limit of experimental viscosity of metallic liquids \cite{nussinov}.

I now ask what constraints are imposed on the fundamental constants from essential life processes in and between cells where motion and flow are involved. Let's write the Navier-Stokes equation as

\begin{equation}
\rho\frac{\partial {\bf v}}{\partial t}=-\grad p+\eta\nabla^2{\bf v}
\label{navier}
\end{equation}

\noindent where ${\bf v}$ is the fluid velocity which is assumed to be small, $p$ is pressure, $\rho$ is density and $\eta$ is dynamic viscosity.

For time-dependent non-equilibrium flow, the solution of Eq. \eqref{navier} depends on kinematic viscosity $\nu=\frac{\eta}{\rho}$. For steady flow, flow velocity depends on $\eta$. The minimum of $\eta$, $\eta_{min}$, can be evaluated as $\eta_{min}=\nu_{min}\rho$, where density $\rho\propto\frac{m}{a_{\rm B}^3}$ and $a_{\rm B}$ is Bohr radius

\begin{equation}
a_{\rm B}=\frac{4\pi\epsilon_0\hbar^2}{m_e e^2}
\label{bohr}
\end{equation}

\noindent where $e$ and $m_e$ are electron charge and mass.

Assuming $m=Am_p$, where $A$ is atomic number and setting $A=1$ for the purpose of the following discussion, this gives

\begin{equation}
\eta_{min}\propto\frac{e^6}{\hbar^5}\sqrt{m_pm_e^5}
\label{etamin}
\end{equation}

A useful property related to viscosity is the liquid relaxation time, the average time it takes a molecule to jump from one quasi-equilibrium place to the next. Its minimal value, $\tau_{min}$, gives a characteristic ``elementary'' time associated with molecular motion. $\tau_{min}$ can be evaluated using the Maxwell relation as $\tau_{min}=\frac{\eta_{min}}{G}$ \cite{frenkel} where $G$ is high-frequency shear modulus. Using Eq. \eqref{etamin}, recalling that the upper bound of elastic moduli are set by fundamental constants as $G\propto\frac{E_{\rm R}}{a_{\rm B}^3}$ \cite{myreview}, where $E_{\rm R}$ is the Rydberg energy

\begin{equation}
E_{\rm R}=\frac{m_ee^4}{32\pi^2\epsilon_0^2\hbar^2},
\label{rydberg}
\end{equation}

\noindent gives

\begin{equation}
\tau_{min}\propto\frac{\hbar^3}{m_ee^4}\left(\frac{m_p}{m_e}\right)^{\frac{1}{2}}
\label{tau3}
\end{equation}

$\tau_{min}$ is related to the shortest time scale in the system set by the Debye vibration period, $\tau_{\rm D}$. Writing $\tau_{\rm D}=\frac{1}{\omega_{\rm D}}$, where $\omega_{\rm D}$ is Debye frequency, recalling $\hbar\omega_{\rm D}=E\left(\frac{m_e}{m}\right)^{\frac{1}{2}}$ \cite{sciadv1}, where $E$ is cohesive energy, setting $a$ and $E$ to their characteristic scales $a_{\rm B}$ and $E_{\rm R}$ and using $m=Am_p$ with $A=1$ as before gives $\tau_{min}=\tau_{\rm D}$ up to a constant.

As compared to $\eta_{min}$ in Eq. \eqref{etamin}, $\tau_{min}$ depends on fundamental constants differently: for example, smaller $\hbar$ increases $\eta_{min}$ but decreases $\tau_{min}$. Physically, this is because smaller $\hbar$ gives larger $E_{\rm R}$, increasing the bond energy and bond stiffness. This increases $\omega_{\rm D}$ and decreases $\tau_{\rm D}$. As a result, $\tau_{min}\propto\tau_{\rm D}$ decreases. Hence, $\tau_{min}$, which sets time scale of short-time dynamics, becomes faster in response to the variation of fundamental constants which increases $E_{\rm R}$. On the other hand, viscosity and its minimal value $\eta_{min}$ increase with $E_{\rm R}$ and with the variation of fundamental constants causing this increase: for example, smaller $\hbar$ increases both $E_{\rm R}$ in Eq. \eqref{rydberg} and $\eta_{min}$ in Eq. \eqref{etamin}. I will revisit this point below.

$\eta_{min}$ in Eq. \eqref{etamin} corresponds to maximal diffusion constant $D$ as follows from the Stokes-Einstein relation $D=\frac{k_{\rm B}T}{6\pi r\eta}$, where $r$ is the radius of a moving particle. This gives

\begin{equation}
D_{max}\propto\frac{1}{\eta_{m}}\propto\frac{\hbar^5}{e^6}\frac{1}{\sqrt{m_pm_e^5}}
\label{dmax}
\end{equation}

Eqs. \eqref{nu1}, \eqref{etamin}, \eqref{tau3} and \eqref{dmax} set the limits for important properties governing dynamics, motion and flow in terms of fundamental physical constants. I now flip the question and ask what happens to these properties if the fundamental constants were different?

$\eta_{min}$ in Eq. \eqref{etamin} and $D_{max}$ in \eqref{dmax} are quite sensitive to $\hbar$ and $e$. If the viscosity minimum $\eta_{min}$ increases due to, for example, smaller $\hbar$ or larger $e$, viscosity necessarily becomes higher at {\it all} conditions of pressure and temperature, in {\it all} liquids (and not just in water essential in our world). This is seen in Fig. 1. At the same time, diffusion decreases according to Eq. \eqref{dmax}, slowing down all diffusive processes of essential substances in and across cells. Physically, the origin of this slowing down due to smaller $\hbar$ or larger $e$ is related to the decrease of the Bohr radius \eqref{bohr}. This results in the increase of the cohesive energy $E=\frac{\hbar^2}{2m_ea_{\rm B}^2}$, which is the Rydberg energy \eqref{rydberg}. The increase of cohesive energy makes it harder to flow and diffuse because a flow event requires overcoming the energy barrier set by the cohesive energy.

Higher viscosity means that water flows slower, dramatically affecting vital flow processes in and between cells and so on. Large viscosity increase (think of viscosity of tar and higher) means that life might not exist in its current form or not exist at all. Consider, for example, blood viscosity: its normal range is about (3.5-5.5) cP. Were viscosity to move significantly outside this range, body functions would be disabled. Changing $\hbar$ or $e$ in Eq. \eqref{etamin} by a few per cent only already covers the normal range and precludes significantly larger variations of these constants. Higher viscosity also slows down essential chemical reactions involved in life processes such proteins folding and enzyme kinetics \cite{chemrate1,chemrate2,chemrate3}.

Once might ask if viscosity increase due to different fundamental constants may be part of the overall slowing down (similar to a video slow motion) whereby all processes slow down but remain functioning. Several observations can be made in this regard. First, larger viscosity not only slows down dynamics but can arrest a life process. Examples include a transition corresponding to the explosive increase of the coagulation rate in biological fluids such as protein solutions and blood. This takes place at the critical value of the P\'eclet number which depends on viscosity \cite{zaccone-peclet}. Second, $\eta_{min}$ in Eq. \eqref{etamin} increases with $e$, $m_e$ and decreases with $\hbar$, whereas the elementary time $\tau_{min}$ in Eq. \eqref{tau3}, or the shortest time $\tau_{\rm D}$, do the opposite. As $\eta_{min}$ and viscosity increase due to, for example, larger $e$ or $m_e$ or smaller $\hbar$, $\tau_{min}$ decreases. This implies that in terms of the shortest atomic timescale $\tau_{min}$ ($\tau_{\rm D}$), time effectively runs faster, and processes dependent on short-time dynamics speed up rather than slow down. Third, the chemical reaction rates of vital biological processes involving, for example, dynamics of proteins and enzymes, $k$, vary as $k\propto\frac{1}{\eta^n}$, where $n$ varies in quite a large range: from 0.3 \cite{chemrate1} to 2.4 depending on the reaction (see, e.g., Ref. \cite{chemrate3} for review). Therefore, viscosity increase affects different reaction rates differently and disrupts the existing balance between products of different reactions and important interactions between those products. Depending on the degree and nature of this disruption, the result can either be finding a new functioning sustainable balance during life development and hence a different type of life or not finding a sustainable living state at all.

One might also ask if cellular life could find a hotter place where overly-viscous and bio-unfriendly water is thinned. This would not work: $\eta_{min}$ and $\nu_{min}$ set the minimum below which viscosity can not fall regardless of temperature or pressure (see Fig. 1). This applies to {\it any} liquid and not just water and therefore to all life forms using the liquid state to function.

%One might also ask if viscosity increase due to different fundamental constants may be part of the overall slowing down (similar to a video slow motion) whereby all processes, including chemical reactions, slow down but remain functioning. However, the elementary time $\tau_{min}$ in Eq. \eqref{tau3}, or the shortest time $\tau_{\rm D}$, depends on fundamental constants differently from $\eta_{min}$ in Eq. \eqref{etamin}: $\eta_{min}$ increases with $e$, $m_e$ and decreases with $\hbar$, whereas $\tau_{min}$ does the opposite. As viscosity increases due to, for example, larger $e$ or $m_e$ or smaller $\hbar$, or the appropriate variation of all three constants, $\tau_{min}$ decreases. This implies that in terms of $\tau_{min}$ ($\tau_{\rm D}$), time effectively runs faster, and processes dependent on atomic dynamics such as chemical reactions speed up rather than slow down.
%The characteristic electron time, in view of Eq. \eqref{rydberg}.

\subsection{Bio-friendly window}

Let $\eta_0$ and $\nu_0$ be viscosities above which life processes are disabled and $D_0$ be the diffusion constant below which life processes are disabled. Conditions for viscosity and diffusion to be bio-friendly are

\begin{eqnarray}
\begin{split}
& \eta_{min}<\eta_0\\
& \nu_{min}<\nu_0\\
& D_{max}>D_0
\end{split}
\label{condi1}
\end{eqnarray}

Each property, $\eta$, $\nu$ and $D$ acts in different life processes and can therefore disable them independently. $\eta$ sets steady flow under external pressure gradient such as active transport or flow in a biochemical machine discussed below. For time-dependent non-equilibrium flow such as pulsed blood flow, the kinematic viscosity $\nu=\frac{\eta}{\rho}$ in Eq. \eqref{navier} becomes important. Essential diffusive processes such as passive and facilitated transport across cellular and intercellular membranes are set by $D$.

Ascertaining the values of $\eta_0$, $\nu_0$ and $D_0$ requires an input from biochemistry and biology. Here, I pose the question of what $\eta_0$, $\nu_0$ and $D_0$ are for such an inter-disciplinary research. Other questions are: which life processes are most sensitive to changes of $\eta$, $\nu$ and $D$ and are disabled first, at each stage of life development? How does a function slow down at higher $\eta$ and lower $D$ and what is the nature of a living-to-non-living {\it transition} at high $\eta$ or low $D$ at each stage? What is the effect on other dependent processes? Can we envisage other life forms where these effects are different? Regardless of implications for fundamental constants, these questions are probably interesting in life sciences on their own.

Regardless of what $\eta_0$, $\nu_0$ and $D_0$ are, interesting qualitative insights emerge. Combining \eqref{condi1} with Eqs. \eqref{etamin}, \eqref{dmax} and \eqref{nu1} gives

\begin{eqnarray}
\begin{split}
& \frac{e^6}{\hbar^5}\sqrt{m_pm_e^5}<\eta_0\\
& \frac{\hbar^5}{e^6}\frac{1}{\sqrt{m_pm_e^5}}>D_0\\
& \frac{\hbar}{\sqrt{m_em_p}}<\nu_0
\end{split}
\label{condi3}
\end{eqnarray}

\noindent where I skipped numerical factors unimportant to establishing the range of variation of each constant in response to the bio-friendly range of $\eta_0$, $D_0$ and $\nu_0$.

Inequalities \eqref{condi3} show how each constant is constrained provided other constants do not vary:

\begin{eqnarray}
\begin{split}
& {\rm max}\left(\frac{1}{\eta_0^\frac{1}{5}}, D_0^{\frac{1}{5}}\right)<\hbar<\nu_0\\
& \frac{1}{\nu_0^2}<m_e<{\rm min}\left(\eta_0^{\frac{2}{5}},\frac{1}{D_0^\frac{2}{5}}\right)\\
& \frac{1}{\nu_0^2}<m_p<{\rm min}\left(\eta_0^2,\frac{1}{D_0^2}\right)\\
& e<{\rm min}\left(\eta_0^{\frac{1}{6}},\frac{1}{D_0^\frac{1}{6}}\right)\\
\end{split}
\label{condi2}
\end{eqnarray}

In \eqref{condi2}, I dropped the conversion factors which can be reinstated using previous equations, for \eqref{condi2} serves to show a trend. As mentioned earlier, the conditions for $\eta_{min}$ and $D_{max}$ are independent because they come from different processes. Hence I used the maximum for the constraint on $\hbar$ and the minimum for constraints on $m_e$, $m_p$ and $e$ in \eqref{condi3} so that the range \eqref{condi2} reflects the mechanism which disables a life process first.

An interesting observation from \eqref{condi2} is that bio-friendly constraints on $\eta$, $D$ and $\nu$ imply a bio-friendly {\it window} for $\hbar$, $m_e$ and $m_p$. This is because $\eta_{min}$ and $\nu_{min}$ depend on $\hbar$, $m_e$ and $m_p$ in \eqref{condi3} differently.

I have considered viscosity getting too high and bio-unfriendly due to different fundamental constants increasing the lower viscosity bound. We could also consider changing fundamental constants to reduce viscosity and its lower bound. If viscosity becomes too low and flow and diffusion get too fast, accumulation of chemicals in cells and organisms can become too large for healthy functions. However, healthy viscosity and diffusion can be recovered by finding different external conditions serving to increase viscosity (see Fig. 1) and decrease diffusion back to their healthy levels if needed. Hence decreasing the lower viscosity bound is not as arresting for life as increasing the lower bound.

\subsection{Fundamental velocity gradient}

To complete the discussion of the role of fundamental constants in life processes involving motion and flow, I derive the fundamental velocity gradient that can be set up in biochemical machines (molecular, cellular, inter-cellular or other). These machines play a vital role in sustaining cells and life. Let's consider a machine creating an external force acting to move the liquid in or between cells. There is a limit to how efficient these machines are because they are powered by chemical energy, the energy of chemical bonds with a characteristic scale set by Eq. \eqref{rydberg}. Let's consider a liquid flowing with constant speed $u$ in direction $x$ in a volume $V$. The viscous stress is $\sigma_x=\eta\frac{\partial u}{\partial y}$, where $y$ is perpendicular to $x$ and $v=\frac{yu}{l}$ in a simple planar geometry, where $l$ is distance between planes \cite{landaufluid}. The viscous force is $f_x=\eta\frac{u}{l}S$, where $S$ is the area across our volume. The work to move the liquid distance $x$ is then $A=\eta\frac{u}{l}Sx=\eta\frac{u}{l}V$. The energy to do this work comes from released chemical, cohesive, energy $E$ (e.g., in the Krebs cycle in the metabolic flux) so we write $\frac{E}{V}=\eta\frac{u}{l}$. $E$ can be written as $NE_0$, where $N$ is the number of energy-releasing centres in a chemical network and $E_0$ is the cohesive energy in one bond whose order of magnitude is given by $E_{\rm R}$ in Eq. \eqref{rydberg}. $V$ can be written as $NV_0$, where $V_0$ is the elementary volume approximately given by $a_{\rm B}^3$. This gives

\begin{equation}
\eta\frac{u}{l}=C\frac{E_{\rm R}}{a_{\rm B}^3}
\label{machine}
\end{equation}

\noindent where the coefficient $C$ absorbs different factors such as the density of energy-releasing centres, their energy and size in relation to $E_{\rm R}$ and $a_{\rm B}$, geometry of the molecular or cellular machine and so on. $C$ is expected to be $C\ll 1$ since $E_{\rm R}$ is larger than a typical energy released in one event in the metabolic flux and $a_{\rm B}$ is smaller than a typical size of the energy-releasing centre.

%Eq. \eqref{machine} relates viscosity and the velocity gradient $\frac{u}{l}$ that can be set up in a liquid by a machine powered by chemical energy. $\frac{E_{\rm R}}{a_{\rm B}^3}$ in Eq. \eqref{machine} is fixed by fundamental physical constants and gives the maximal efficiency of a molecular or cellular machine to move the viscous liquid.

I now recall the lower bound for viscosity discussed earlier, $\eta_{min}<\eta$. Combining this inequality with Eq. \eqref{machine} gives

\begin{equation}
\eta_{min}\frac{u}{l}<C\frac{E_{\rm R}}{a_{\rm B}^3}
\label{machine2}
\end{equation}

Writing $\eta_m=\nu_m\rho$, $\rho=\frac{m}{a_{\rm B}^3}$, $m=Am_p$ and using Eq. \eqref{nu1} and Eq. \eqref{rydberg} as before, I find

\begin{eqnarray}
\begin{split}
& \frac{u}{l}<\left(\frac{u}{l}\right)_{max}\\
& \left(\frac{u}{l}\right)_{max}=\frac{C}{8\pi\sqrt{A}\epsilon_0^2}\frac{{m_e}e^4}{\hbar^3}\left(\frac{m_e}{m_p}\right)^{\frac{1}{2}}
\end{split}
\label{machine3}
\end{eqnarray}

Eq. \eqref{machine3} gives the upper bound for the velocity gradient that can be set up by a biochemical machine powered by the chemical bond energy in terms of fundamental physical constants.

Using Eq. \eqref{machine3}, I introduce  the fundamental velocity gradient set by fundamental constants as:

\begin{equation}
\left(\frac{u}{l}\right)_f\propto\frac{{m_e}e^4}{\hbar^3}\left(\frac{m_e}{m_p}\right)^{\frac{1}{2}}
\label{machine4}
\end{equation}

\noindent with the dimensionality inverse of $\tau_{min}$ in \eqref{tau3}.

\subsection{Variability at fixed $\alpha$ and $\beta$}

I have considered how bounds on viscosity, flow, diffusion and velocity gradient can change in response to varying fundamental constants. This variation should be constrained because it should avoid the range where the production of lower-level structure, such as atoms, is disabled. In particular, the fine-structure constant $\alpha=\frac{e^2}{\hbar c}$ and the proton-to-electron mass ratio $\beta=\frac{m_p}{m_e}$ are considered finely-tuned in order for heavy nuclei to be produced in stars \cite{barrow,carrbook,finebook}. Hence I fix $\alpha$ and $\beta$ to reflect this tuning and ask how this affects viscosity, diffusion and flow. One way to write $\eta_{min}$ in Eq. \eqref{etamin}, $D_{max}$ in Eq. \eqref{dmax}, $\nu_{min}$ in Eq. \eqref{nu1} and $\left(\frac{u}{l}\right)_f$ in Eq. \eqref{machine4} in terms of $\alpha$ and $\beta$ is

\begin{eqnarray}
\begin{split}
& \eta_{min}\propto\left(\frac{e^2}{\hbar c}\right)^3\sqrt{\frac{m_p}{m_e}}\left(\frac{m_ec}{\hbar}\right)^3\hbar\\
& D_{max}\propto\frac{1}{\left(\frac{e^2}{\hbar c}\right)^3\sqrt{\frac{m_p}{m_e}}}\left(\frac{\hbar}{m_ec}\right)^3\frac{1}{\hbar}\\
& \nu_{min}\propto\frac{1}{\frac{e^2}{\hbar c}\sqrt{\frac{m_p}{m_e}}}\frac{e^2}{m_ec}\\
& \left(\frac{u}{l}\right)_f\propto\frac{\left(\frac{e^2}{\hbar c}\right)^2}{\sqrt{\frac{m_p}{m_e}}}\frac{m_ec^2}{\hbar}
\end{split}
\label{fixed}
\end{eqnarray}

I note that bounds \eqref{fixed} derived in non-relativistic condensed matter physics are not expressible in terms of $\alpha$ only, as is often the case for other bounds and properties in relativistic high-energy physics \cite{barrow,barrow1,carrbook}, but depend on other fundamental constants too.

%in passing that the inverse of $\frac{m_ec}{\hbar}$ in $\eta_{min}$ is the reduced Compton wavelength of the electron and $\frac{m_ec^2}{\hbar}$ in $\left(\frac{u}{l}\right)_f$ is the Compton frequency of the electron.

Fixing $\alpha$ and $\beta$ still leaves many ways of varying $\eta_{min}$, $D_{max}$ and $\nu_{min}$. For example, any change of $\hbar$, $m_e$ or $c$ in the factor $\frac{m_e^3c^3}{\hbar^2}$ in $\eta_{min}$ and $D_{max}$ changes $\eta_{min}$ and $D_{max}$ but this change can always be compensated by changing other constants in $\alpha$ and $\beta$ to keep $\alpha$ and $\beta$ intact. This can be done in many ways: changing $m_e$ can be compensated by $m_p$ to keep $\beta$ intact; changing $\hbar$ in $\frac{m_e^3c^3}{\hbar^2}$ can be compensated by $e$ to keep $\alpha=\frac{e^2}{\hbar c}$ intact, and so on. Similarly, changing $m_e$ in the factor $\frac{e^2}{m_ec}$ in $\nu_{min}$ changes $\nu_{min}$ but can be compensated by $m_p$ to keep $\beta$ intact. Or, changing $e$ in the factor $\frac{e^2}{m_ec}$ changes $\nu_{min}$ but can be compensated by the change of $\hbar$ in $\alpha$, and so on. The fundamental gradient can also be varied in ways which keep $\alpha$ and $\beta$ intact.

%Although the Compton wavelength or frequency are physically unrelated to viscosity, Eq. \eqref{fixed} shows that viscosity and diffusion bounds can still vary even if we incorporate constraints from quantum and relativistic physics involved in Compton scattering and fix $\lambda_{\rm C}$, together with fixing $\alpha$ and $\beta$ (for example, we can change $m_e$ and $\hbar$ proportionately in $\eta_{min}\propto\lambda^3_{\rm C}\hbar$ in \eqref{fixed} so that $\eta_{min}$ is changed but $\lambda_{\rm C}$ is intact and then adjust $m_p$ to keep $\beta$ intact and change $e$ in $\alpha$ to keep $\alpha$ intact, and so on). Differently from $\eta_{min}$, fixing the Compton frequency (together with $\alpha$ and $\beta$) fixes the fundamental gradient $\left(\frac{u}{l}\right)_f$.

We therefore see that a Universe with fundamental constants different from ours can produce heavy elements in stars but have a planet where all liquids have very high viscosity due to large $\eta_{min}$ in \eqref{fixed}, for example that of tar or higher and where observers may not emerge. This can be achieved, for example, by increasing $m_e$ and/or decreasing $\hbar$ while keeping $\alpha$ and $\beta$ constant in \eqref{fixed} as discussed above. In order to reduce this high life-disabling $\eta_{min}$ to its current bio-friendly value, we need to dial the fundamental constants back to their current values so that the bounds \eqref{fixed} become bio-friendly. Hence we need to tune the same fundamental constants setting $\alpha$ and $\beta$ ($\hbar$, $e$, $c$, $m_e$, $m_p$) which, importantly, involves tuning which is additional and different to tuning involved in fixing $\alpha$ and $\beta$. This additional tuning due to bio-friendly viscosity is not needed for the generation of heavy nuclei and is therefore {\it redundant} for heavy nuclei. This redundancy involves vast, up to 15 orders of magnitude, differences between the two processes (generating heavy nuclei and bio-friendly viscosity and diffusion in living organisms) in terms of size and similarly large differences in terms of energy.

The above redundancy applies if tight constraints on $\alpha$ are relaxed \cite{adamsreview}. In this more general case, the constraints on the fundamental constants from bio-friendly viscosity and diffusion in \eqref{fixed} are still additional and different from those imposed by the production of heavy nuclei.

%I note that I arrived at this redundancy for reasons unrelated to the need for observers to exist as posited in SAP. Rather, I (a) observed that the current values of fundamental constants are conducive to bio-friendly viscosity and diffusion involved in essential life processes and (b) combined this observation with the result that viscosity and diffusion have bounds set by fundamental constants.

These observations bear a relation to questions asked previously: can we understand the values of fundamental constants on the basis of a theory more fundamental than we currently have (the Standard Model) \cite{barrow,barrow1,carrbook,finebook,weinberg}? How were these constants tuned \cite{carrbook,finebook}? One possibility is that fundamental constants were tuned (attained their currently observed values) once. As mentioned earlier, this would involve redundancy.

If redundancy is to be avoided, we can conjecture that multiple independent tunings were involved. This includes tuning fundamental constants to produce heavy nuclei and additional tunings needed for other observed sustainable structures to emerge. This conjecture of multiple tuning suggests a similarity to biological evolution where functionally similar traits, such as the different optical nerve connections in humans and octopi, were acquired independently. If the analogy is between acquiring a new trait and one act of tuning fundamental constants leading to a new set of these constants, then an organism as a system with multiple separately acquired traits is analogous to the set of observed fundamental constants produced as a result of multiple tunings. The evolutionary mechanism changes the focus of discussion of fundamental constants and their values: currently observed constants can be considered analogous to, for example, a set of traits acquired by the human eye as the result of past evolutionary changes. These traits were helpful and stayed but the question as to the number of traits in this set is not viewed as meaningful. Nor is this set most optimal (a human eye is less optimised as compared to the octopus eye), as is the case with our Universe which could have been more habitable were some fundamental constants different \cite{adamsreview}. An analogy with physics would imply that the observed fundamental constants are the result of nature arriving at sustainable physical structures, but the values of these constants may not need to be derived in a more fundamental theory as considered previously \cite{barrow,barrow1,carrbook,finebook,weinberg}.

A useful example of such an emergence of sustainable structures in biochemistry are the DNA blocks forming in protocells as a result of positive feedback in the metabolic flux \cite{lanebook}. This positive feedback is not just a general idea but is based on specific biochemical reactions: the core metabolism central to life probably started when first catalysts sped up helpful aspects of the metabolic flux in protocells, enabling the conversion of H$_2$ and CO$_2$ into the fabric of new protocells. The first nucleotides, followed by RNA and DNA, then emerged inside such replicating protocells through the positive feedback: protocells with more beneficial chemicals replicated better and passed these chemicals to their daughter cells. This led to the insight that ``meaning emerged with function'': the DNA - the mathematical structure - emerged in the process of helping protocells get better at copying themselves. This enabled protocells to proliferate and hence sustain themselves \cite{lanebook}. I discuss this in more detail elsewhere.

%\section{Conclusions}

In summary, I showed how condensed matter physics and liquid physics in particular provides insights into fundamental constants in addition to those discussed in high-energy physics. The fundamental constants have a window constrained by bio-friendly viscosity and diffusion. Ascertaining this window quantitatively invites another inter-disciplinary input from life sciences and raises new questions such as the living-to-non-living transition as a function of viscosity and diffusion. Once ascertained, we can compare the bio-friendly window with constraints on fundamental constants from high-energy physics \cite{barrow,barrow1,carr,cahnreview,hoganreview,adamsreview,uzanreview,carrbook,finebook}. We saw that bounds on viscosity, diffusion and the fundamental velocity gradient in a biochemical machine can all be varied with no implication for the production of heavy nuclei. %This leads to a conjecture of multiple tuning and an evolutionary mechanism.

Acknowledgements: I am grateful to F. Adams, S. Arseniyadis, V. V. Brazhkin, B. Carr, R. Goldstein, S. de Kort, L. Noirez, N. Ojkic, Isabel M. Palacios, A. E. Phillips, U. Windberger and A. Zaccone for discussions and to the EPSRC for support.

%Competing interests: The author declares no competing interests.

%Data and materials availability: All data needed to evaluate the conclusions in the paper are present in the paper and references.

%\bibliography{refs}

\bibliographystyle{apsrev4-1}

\end{document}